# Economic Consequences of the COVID-19 Pandemic on Sub-Saharan Africa: A historical perspective

**Anthony Enisan AKINLO[1] and Segun Michael OJO[2]**

[1]Department of Economics, Obafemi Awolowo University, Ile-Ife
Email: aakinlo@oauife.edu.ng

[2]Department of Economics, Redeemers University, Ede
Email: ojosegunm@yahoo.com

**Abstract**

This paper examined the economic consequences of the COVID-19 pandemic on sub-Saharan Africa (SSA) using the historical approach and analysing the policy responses of the region to past crises and their economic consequences. The study employed the manufacturing-value-added share of GDP as a performance indicator. The analysis shows that the wrong policy interventions to past crises led the sub-Saharan African sub-region into its deplorable economic situation. The study observed that the region leapfrogged prematurely to import substitution, export promotion, and global value chains. Based on these experiences, the region should adopt a gradual approach in responding to the COVID-19 economic consequences. The sub-region should first address relevant areas of sustainability, including proactive investment in research and development to develop homegrown technology, upgrade essential infrastructural facilities, develop security infrastructure, and strengthen the financial sector.

**Introduction**

The outbreak of the the COVID-19[1] pandemic is relatively unique both in presentation and in magnitude (Boissay and Rungcharoenkitkul, 2020). The economic consequences of the COVID-19 pandemic are incomparable to those of the human immunodeficiency virus (HIV) and acquired immunodeficiency syndrome (AIDS), and the Ebola outbreak of 1976 (CDC, 2016). According to the International Monetary Fund (IMF, 2020), the COVID-19 pandemic is the worst global crisis since the Great Depression of the 1930s. The HIV/AIDS and Ebola outbreaks did not result in global-crisis, lockdown, and paralyzing of economic activities in virtually all countries simultaneously as was the case with the COVID-19 pandemic. Other crises which had similar effects as COVID-19 were the First World War, The Great Depression of the 1930s, and the Second World War. Although these were not disease-related, they had the same magnitude of impact as COVID-19. COVID-19 is not a sectional or regional crisis, it is a pandemic of global magnitude (Gennaro et al., 2020). The supply and demand-side shocks that accompanied the COVID-19 pandemic are unprecedented in the world's recent economic history. This development is likely to bring about a change in the world economic order as happened in the wake of the First World War, the Great Depression, and the Second World War.

The economic disruptions that have accompanied the COVID-19 pandemic involve diverting countries' national expenditures from other sectors to health, food, and security. In other words, the COVID-19 pandemic brought about a sudden diversion of government spending to the medical and paramedical sectors, food supply, and security infrastructure due to lockdown operations. Some positive developments from the COVID-19 pandemic include improved medical facilities, enhanced food supply capacity, and improved security operatives in many nations. These outcomes are expected to impact positively on the economies of the countries. In contrast, the havoc the pandemic wreaked is unquantifiable in terms of loss of lives, unplanned medical and burial expenditures, unemployment burden, business sector revenue loss, and the general disruption of economic activities, outweigh its positive features. Consequently, a global economic recession may be inevitable.

---

[1] Coronavirus disease 2019 (commonly called COVID-19) is referred to as an infectious disease caused by a novel coronavirus also known as severe acute respiratory syndrome coronavirus 2 (SARS-CoV-2; formerly called 2019-nCoV). It was first noticed in connection with an outbreak of respiratory illness cases in Wuhan City, Hubei Province, China.

**Historical Briefing on the World Economy and Global Crises (Mid-18th Century till Date)**

The well-organized present world economic system did not start in the manner it operates today. The world economy has gone through many evolutionary developments that culminated in the current economic system. In other words, the current economic system and the current world economic order are products of historical events. The successes and failures that trailed the responses of nations and groups of countries to the aftereffects of some past notable crises brought the world to its present position. Therefore, SSA can learn from history on how to carry out proactive post-COVID-19 economic recovery plans. The failure of SSA countries to respond appropriately to crises has severally thrown the region into stagnation and perpetual low economic performance. It is therefore high time that the sub-region learnt from history to avoid total collapse of their economies.

The current international economic order is rooted in the economic revolution of the mid-18th century. Before this period, the economies of most nations operated at subsistence level. In other words, the world economy was significantly enhanced by the breakthrough in communication and transportation infrastructures of the 18th century, which boosted the exchange of goods and services among the nations of the world (Shafaeddin, 1998). The economic bonds that ensued from the resultant economic interdependence significantly enhanced the efficient utilization of the world's spatially distributed economic and natural resources. This development helped to instill the practice of the comparative advantage paradigm in the international economic scene for the first time. Other scientific and technological discoveries in the late 18th century and early 19th century laid the foundation for Britain's Industrial Revolution of the 19th century, which pioneered and motivated industrialization in other industrialized nations.

Before the outbreak of the First World War, the world economy operated on the gold standard and its attendant international trade. The first global crisis that truncated the first world economic order under the gold standard was the First World War. The war started at a time when the world economic system was relatively stable with smooth flows of transactions across the nations' boundaries. Put differently, the gold standard worked well until the outbreak of the First World War in 1914. However, the gold standard mechanism could not control the economic complications that arose on the international economic scene as a result of the war. The severity of the war on the global economy was consequential, to the extent that it put the existing world economic order in disarray. The gold standard mechanism collapsed and was abandoned for a free exchange rate system whereby individual countries reverted to their national currencies. The war revealed, among other things, that the cooperation and mutual understanding shared by the trading countries were instrumental to the working of the gold standard mechanism. After the war, the free exchange rate system that replaced the gold standard failed to restore the pre-war world economic order. The failure could be attributed to the free exchange rate system which was characterized by strong competition among the countries and their currencies. However, Britain re-embraced the gold standard and all the colonies under the British Empire followed suit. The British group had a large membership, which assisted them in securing global acceptance of the gold standard for the second time. Afterward, Britain gained economic dominance over other countries, as did the British pound sterling. Therefore, the pound sterling and gold were the dominant currencies, and countries worldwide were holding foreign reserves in gold and pound sterling (Dwivedi, 2009).

The new gold system could not restore the stable pre-war international economic system due to speculations and other measures. However, the British economy ran into a crisis, a development that led to the British losing gold to other countries like France and other European countries. Consequently, the British government withdrew from the gold standard because it had lost its gold. Expectedly, the Commonwealth of Nations followed Britain in abandoning the gold standard, which led to total abandonment of the gold standard in the world economic system. In 1929, the US stock market crashed. The incident led the world into the Great Depression of the 1930s due to inappropriate policy responses. The world was barely recovering from the 1930s Great Depression when the Second World War broke out due to some political complications. In the early 20th century, the world economy was traumatized by three global calamities namely: the First World War, the 1930s Great Depression, and the Second World War.

The post-World War II era was another action-packed phase in world economic history. The period witnessed two geopolitical phenomena: the Cold War and the decolonization of former colonies. The economic confusion that trailed the war also motivated the inauguration of notable international trade and financial institutions, such as the International Bank for Reconstruction and Development (IBRD) or the World Bank, the International Monetary Fund (IMF), and the International Trade Organization (ITO) which later metamorphosed into the World Trade Organization (WTO). These institutions were established to fashion a new economic order that would prevent the reoccurrence of the events that led to the Second World War and the Great Depression. Each of the international institutions has a primary responsibility for which it was created. The International Bank for Reconstruction and Development (IBRD) or the World Bank was established to rebuild the war-battered economies, particularly in Europe and Asia, the two regions that bore the brunt of the war. The International Monetary Fund (IMF) was saddled with maintaining a fixed exchange rate system called the Bretton Woods System. The Bretton Woods exchange rate system collapsed in the early 1970s. The International Trade Organization (ITO) initially did not materialize due to some political factors but later restructured as the World Trade Organization (WTO). The organization assists in liberalizing international trade at the global level.

The developed countries that were adversely affected by the war had a rapid recovery, such that their economies stabilized within the first decade after the war. In contrast, the sub-Saharan African economies were stagnant. The concern over the developmental challenges of the developing countries drew the attention of policymakers and the development partners first post-Second World War. Before this time, the majority of developing countries were colonies of the developed countries which used them as sources of raw materials and cheap labour for industrial activities in their home countries. Since many of the developed countries at their takeoff did rigorous industrialization and manufacturing sector development, the same policy was recommended as a catch-up strategy for developing countries. The advocacy for industrialization and manufacturing output development as a catch-up strategy for the developing countries (Prebisch, 1950; Singer, 1950) was reinforced by the groundbreaking publications of Nicholas Kaldor in the mid-1960s, which came to be known as Kaldor's law. Kaldor's law is in two forms, namely Kaldor's first and Kaldor's second law (or Verdoorn's law) (Kaldor, 1966). The first law says that manufacturing is the engine of growth, while the second law postulates that there is a positive causal relationship between output and labour productivity in manufacturing derived from static and dynamic returns to scale.

The wide acceptance of this theory reinforced the adoption of industrialization and manufacturing output development as a catch-up strategy for developing countries through the import substitution policy (Lin, 2011). Moreover, the adoption of the import substitution policy by the developing countries ensured economic freedom from the colonists. The primary products of the colonies were exported to the colonists' countries for industrial uses, and consequently, the colonists did not promote industrial development or infrastructural upgrading in their colonies. Thus, when the colonies gained their independence the consensus was that they should stop the exportation of their raw materials. This implied that the developing countries (colonies) would utilize their raw materials locally and develop their industrial sectors. Hence, import substitution was adopted to prevent neo-colonization and further resource exploitation.

Throughout the 1960s and 1970s, import substitution was the industrialization policy tool that held sway till the early 1980s, when it was replaced with the export promotion strategy (Bruton, 1998; Bennett, Anyanwu and Alexanda, 2015). The import substitution agenda failed due to low value-added coupled with the market limitation to low-technology commodity production where the African sub-regions' import substitution agenda made little progress (IBRD, 1971). Generally, import substitution is constrained by the size of the domestic market except the products are exportable. Unfortunately, the import substitution products in sub-Saharan Africa were not internationally competitive, hence the collapse of the import substitution scheme in the region (Krueger, 1974; Krugman, 1993; Soludo, Ogbu and Chang 2004).

However, another school of thought emerged in the late 1970s that advocated for foreign direct investment (FDI) as the appropriate means of spurring industrialization and manufacturing sector development through technological transfer. Koizumi and Kopecky (1977) were the first to model FDI and technology transfer.

In 1982, another landmark publication that tremendously reinforced the plausibility and theoretical relevance of the work of Nicholas Kaldor was released by Gavin Kitching, who argued that "if you want to develop you must industrialize" (Kitching, 1982). Kitching premised his argument for industrialization as the catch-up strategy on the limitation of agriculture and the efficiency of economies of scale. He laid the foundation for the structural transformation approach that constitutes the main policy thrust of the late industrialization campaign (UNIDO, 2015). Essentially, Kitching work reinforced the plausibility of the adopting industrialization and manufacturing development as catch-up strategies for developing economies.

As a result, FDI expansion and export promotion targets fostered under the defunct Structural Adjustment Programme (SAP) were the policy measures for industrialization till the collapse of SAP and its inbuilt policy measures in the early 1990s. The region's manufacturing sector was characterized by low value-added firms and the export industries engaged in semi-processing primary products (Ebenyi et al., 2017; Bennett, Anyanwu and Alexanda, 2015). Moreover, it was argued that local firms that lacked the necessary absorptive capacity for advanced technology and skills could not benefit from the FDI knowledge spillover (Blomstrom and Kokko, 2003). Besides, the high wave of globalization and the high intensity of the cross-border trades that followed the technological breakthroughs in the areas of ICT, mobile communication technologies, and online transactions have turned the world into a global village. This development has now opened the way for the global value-chains campaign which argues for the global fragmentation of production activities (Timmer et al., 2014), starting from the early 1990s till date.

**Recap on Sub-Saharan African Industrialization Drive**

This paper focuses on the industrialization performance of SSA as it is central to the economic pursuit of the region. The industrialization scheme is the major policy intervention used by SSA countries to address past economic crises. These include the 1973 Oil Crisis Recession, the 1980 Energy Crisis Recession, the 1981 Iran Energy Crisis Recession, and the 1990 Gulf War Recession. Besides, Goal 9 of the Sustainable Development Goals (SDG) advocates for inclusive and sustainable industrialization as a means to create employment and generate income towards achieving the desired sustainable development.

The pursuit of the industrialization agenda in SSA started with the adoption of the import substitution policy in the 1960s (Mendes, Bertella, and Teixeira, 2014). Import substitution is an industrialization policy that seeks to replace imports with domestic production. However, the question is: Was import substitution the right policy tool for sub-Saharan African countries at that time considering their infrastructural facilities? Unfortunately, based on the advice of the foreign-trained experts, most countries in SSA at independence opted for the import substitution development strategy. The proponents of the import substitution agenda based their argument on the availability of natural resources in the sub-region. However, adequate consideration was not given to machine equipment, energy consumption, transport facility, road access, technical know-how, financial capital, and the institutional arrangements required to foster a successful import substitution agenda.

Figure 1 shows the trend of manufacturing value-added share of GDP in sub-Saharan Africa from 1965 to 1985, which was the period of the import substitution scheme. Coincidentally, the two past major epidemics broke out within the period of the import substitution scheme. The first epidemic was HIV, was started in 1959 in Kinshasa, the Democratic Republic of the Congo. The second was the Ebola virus was first discovered in 1976 in the same Democratic Republic of Congo. The two diseases did not cause serious economic disruptions like COVID-19 because they did not break out like COVID-19. Besides, HIV/AIDS and Ebola did not spread out throughout the period. Consequently, their economic impacts were not too noticeable. This study employs import substitution policy to explain the responses of the sub-Saharan African countries because it was the economic policy adopted by most countries in the SSA sub-region. Besides, the most accessible indicator of economic performance during the period was the manufacturing value-added share of GDP. Therefore, this study adopts manufacturing value-added share of GDP as the performance indicator to explain the economic performance of SSA during this period.

In the late 1960s, when most sub-Saharan African countries embraced the import substitution policy, the countries recorded positive growth. However, the improvement in MVA growth was short-lived due to the failure of the indigenous firms to sustain growth-enhancing productivity in the infant nations. The situation was relatively better during the 1970s when the oil-exporting sub-Saharan African countries had windfall incomes as a result of the oil crises of the 1970s, which favoured the oil-exporting developing countries. At the same time, the oil-importing developing countries were able to secure huge loans for bailouts from the savings made by their oil-exporting counterparts. Thus, the sub-region had little relief from the early to the mid-1970s. In sum, the import substitution policy did not perform well because it was the wrong policy response to the start-up-challenges of the young independent countries in SSA.

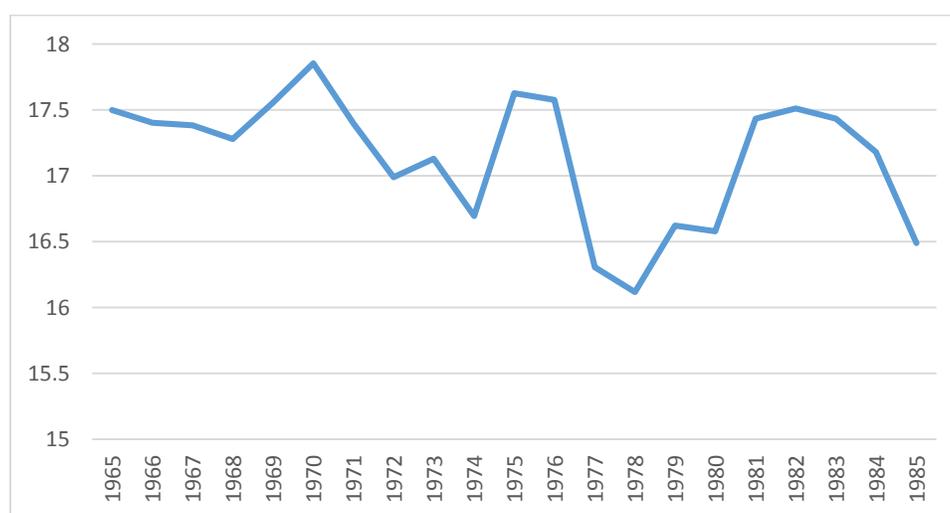

**Figure 1: Manufacturing Value Added Share of GDP (1965-1985)**

*Source:* Author's computation.

The export-promotion drive in sub-Saharan Africa began in the early 1980s, after the two consecutive economic recessions (1980 Energy Crisis Recession and 1981 Iran Energy Crisis Recession) which were the last straw for the import- substitution programme (Johnson and Wilson, 1982). The period also coincided with the expansion of the HIV/AIDS and Ebola epidemics. Though HIV/AIDS and Ebola were horrible diseases they were not ranked as pandemics because their economic impacts were not so severe. Within the first 39 years of the HIV/AIDS outbreak, 75 million people were infected and 32 million people reportedly died from AIDS-related illnesses (PEPFAR, 2020). In West Africa, Ebola claimed about 11,316 lives during the 2014 epidemic. Those epidemics were not drastic like COVID-19, and sub-Saharan African countries managed them alongside other economic and socioeconomic crises. The only noticeable economic consequence of Ebola occurred in Guinea, Liberia, and Sierra Leone. According to a World Bank report, about $2.2 billion was lost in 2015 in three countries. Germany, the USA, and the World Health Organization bore the brunt of the financial burdens of the epidemic for those African countries.

As the HIV/AIDS infection started spreading in the early 1980s, sub-Saharan African countries were implementing the export promotion scheme through the 1980s Structural Adjustment Programme (SAP). The SAP policies included privatization, fiscal austerity, free trade, and deregulation. The economic setback that befell the region under export promotion was more than the failure of import substitution, as evident in Figures 1 and 2. In other words, the manufacturing value-added share of GDP during the import substitution period oscillated but during the export promotion period, it plummeted. The reason for the economic deterioration under the export promotion programme is evident. The unsolved underperformance of the import substitution policy translated into the poor outcome of the export promotion agenda. In an ideal situation, the success of the import substitution policy would lead to the attainment of the export promotion goals and targets.

In Figure 2, the manufacturing value-added curve trend was downward from left to right throughout the export promotion period. Before the adoption of the export promotion agenda, the region's manufacturing value-added as a share of GDP was relatively high. The average performance under import substitution was far higher than under export promotion. In other words, the manufacturing productivity of SSA dissolved into a persistent decline right from the inception of the export promotion scheme.

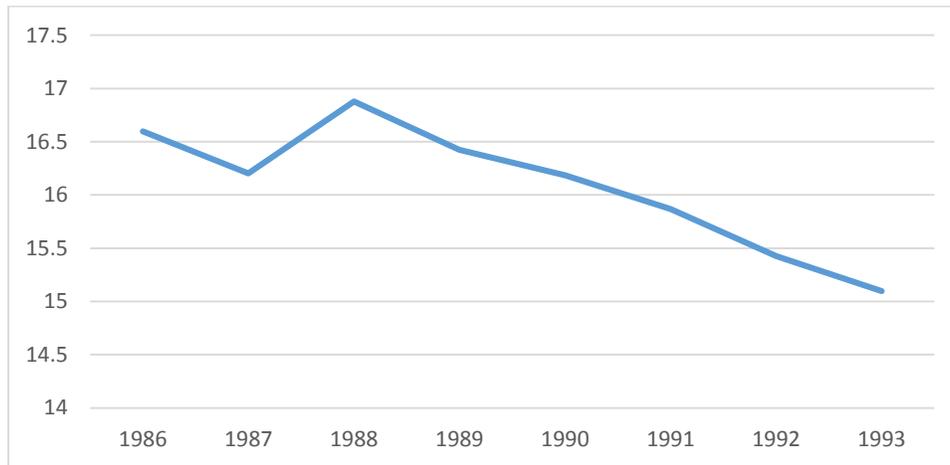

**Figure 2: Manufacturing Value Added Share of GDP (1986-1993)**
*Source:* Author's computation.

Global value chains were the industrialization policy tool that replaced the export promotion agenda in the international manufacturing environment. The scientific and technological breakthrough in information technology in the early 1990s intensified globalization in no small measure. The resultant increase in the volume of trade and interdependence among nations stirred the international fragmentation of the production process. Thus countries tended to specialize in different stages of value addition in the production process. The inability of sub-Saharan African countries to participate competitively in the global value chains was responsible for the continuous decline in manufacturing value-added right since the inception of global value chains trading. It has been argued that any nation or region that fails to participate in the global value chains will be marginalized in the international manufacturing markets (Moris and Fessehaie, 2014). Global value chains have held sway in the international manufacturing market since the early 1990s. Sub-Saharan Africa's manufacturing value-added share of GDP shows that the region's level of participation in global value chains is still low. The deindustrialization trend that had set in since the era of export promotion in the sub-region did not exhibit an iota of improvement till the outbreak of COVID-19. The COVID-19 pandemic is here; where should the sub-region go from here? What is the right policy response for the sub-region?

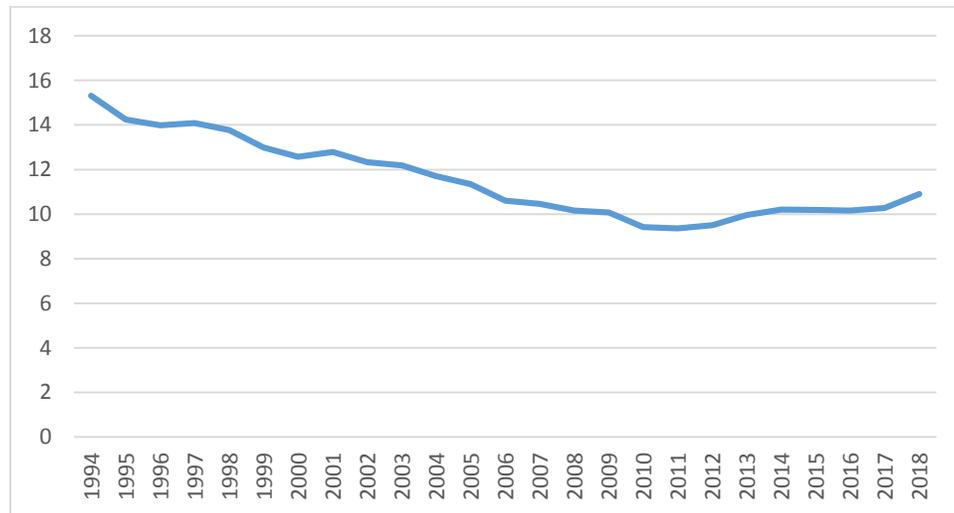

**Figure 3: Manufacturing Value Added Share of GDP (1984-2018)**
*Source:* Author's computation

From the foregoing historical briefing, the current world economy had witnessed three categorical global shocks before the outbreak of the COVID-19 pandemic. The first global crisis was the First World War, followed by the Great Depression of the 1930s and then the Second World War. The economic consequences of these global phenomena were aggravated and prolonged unnecessarily due to inappropriate policy responses. The uncoordinated floating exchange rate system adopted after the First World War and the Bretton Woods policies initiated in response to the economic consequences of the Great Depression and the Second World War only made the SSA countries worse off. This is evident in the persistent economic deterioration that has characterized the region. This development raises the question on the appropriate policy intervention that should be implemented to mitigate the economic consequences of COVID-19 in the sub-region.

The success of import substitution is supposed to facilitate export promotion. Virtually all the developed countries at the initial stage of their industrial development toed the path of import substitution and protection of infant industries through restrictions. Thus, it was after a successful and stable import substitutionthat they engaged in bilateral and multilateral trade agreements to sell their surpluses to the outside world. No developed or industrialized nation ever embarked on export promotion without successful import substitution (Shafaeddin, 1998). The import substitution strategy is still the best approach that a country can use to nurse indigenous firms and thereby build their supply capacity which will enhance the country's competitiveness in export promotion (Shafaeddin, 1998).

Sub-Saharan African countries could not participate well in global value chains due to the inherent barriers in the economic system of the sub-region. Presently, the service sector dominates the economic activities of the sub-region. The average service value-added share of GDP in sub-Saharan Africa over the period 1981 to 2018 was estimated at 47.5%, while the manufacturing value-added share of GDP was 13.1%. A country that fails in export promotion stands a narrow chance of success in global value chain trading. Sub-Saharan African manufacturing value-added share of GDP has declined consistently over the past three decades due to premature deindustrialization. The sub-region moved from failed import substitution to export promotion. It leapfrogged to active industrialization through import substitution instead of utilizing the proceeds from the exports of primary products to build the supply capacity required for sustainable industrialization. Thus, the industrialization experiences of sub-Saharan African countries can be described as a movement from premature industrialization to premature deindustrialization.

Another important observation from the historical review is about international cooperation failure. The world economy has witnessed several instances of international cooperation failures. The first disruption occurred in the wake of the First World War when the gold standard system that held the world economy together lost its grip due to the resultant cessation of the cooperation among nations following the economic hardships that trailed the war. Nations could no longer listen to one another. Instead, every government was preoccupied with internal arrangements and rearrangements amidst speculative and preventive measures against the reoccurrence of the economic hardships caused by the war. The immediate outcome of this scenario was the abandonment of the gold standard, which however later reemerged after the war.

The gold standard system relied on cooperation among countries to operate a fixed exchange rate and international capital mobility. The war pulled countries apart and the economic situation across countries worsened. Individual governments introduced currency and capital controls which put an end to the second era of the gold system. The situation persisted till the outbreak of the Second World War in 1939. The instability in international trade and the deterioration in the world economy caused economic managers to initiate the Bretton Woods Agreements in 1944 to restore cooperation among nations. Historically, whenever an incident of global impact occurs there is the likelihood of abrupt abandonment of existing international cooperation and the emergence of a new world economic order. So, when the dust of the COVID-19 pandemic settles, sub-Saharan African countries should expect a new world economic order because there will be a modification in the existing international economic cooperation, if not a total abandonment. Therefore nations and regions that depend on international agreements and cooperation may be let down amidst the likely economic confusion that may arise in the post-COVID-19 era.

**Economic Consequences of the COVID-19 Pandemic**

The economic consequences of the pandemic on SSA can be viewed from three major perspectives. The first is the present and future impacts of the demand and supply shocks that resulted from the internal lockdown operations on the economic lives of SSA countries. The second is the spillover effects of the collapse of global economic activities on productivity, and the financial stability of SSA as a result of international lockdowns. The third scenario is the fall in commodity prices (particularly oil) that occurred due to the pandemic, which might aggravate the economic hardship of the oil-exporting countries in the region. Figure 4 depicts the transmission channels of the economic consequences of the COVID-19 pandemic in the context of the sub-Saharan African economy. On the one hand, the internal lockdowns in the countries triggered supply and demand shocks through the sudden disruptions inflicted on economic activities. The lockdown disrupted the production of goods and services, thereby leading to supply failure. The resultant revenue reduction engendered unplanned mass downsizing among firms due to limited revenue. This led to increased level of unemployment in the sub-region. One of the immediate effects of the sudden rise in unemployment was a sharp fall in per-capita income because all the sources for generating income were cut off. The shortage in supply of goods and services caused prices of goods and services to rise, which affected both demand and consumption.

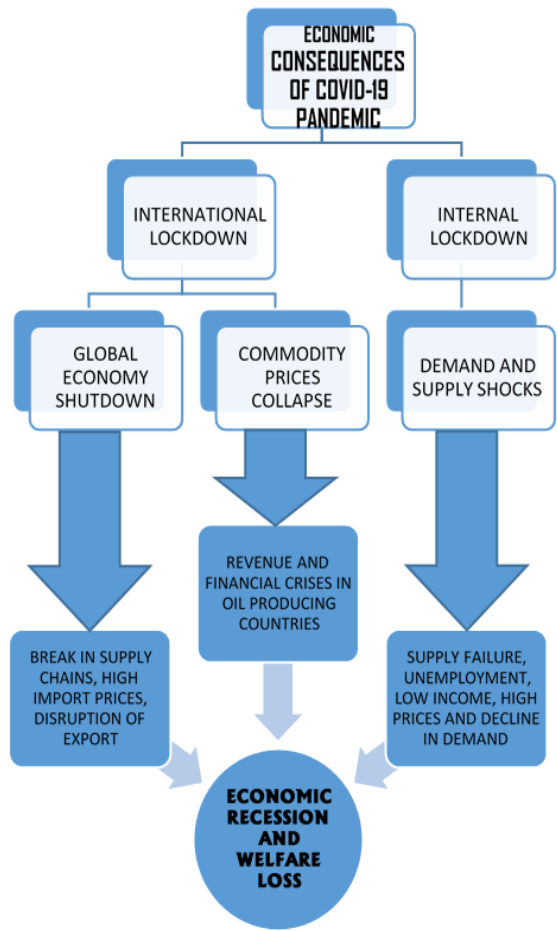

**Figure 4: COVID 19 Economic Consequences Impact Channels.**
*Source:* Author's computation.

At the global level, international lockdowns have adversely affected economic transactions and suppressed commodity prices. The lockdown has led to reduction in manufacturing production and major disruptions in global value chains and products' supply. The break-up of the international supply chain has precipitated high import prices, thereby worsening the fragile economic situation in import-dependent economies like SSA. Many SSA countries get their public revenue and foreign exchange earnings from the export of one primary product or the other. The abrupt international lockdown on exports and the global oil price collapse have thrown many SSA countries into severe financial disaster.

One critical lesson from the COVID-19 pandemic is the need for self-reliance among countries in the SSA sub-region. Whenever there is international cooperation failure, a nation's level of self-reliance determines its level of survival. If a section or region is in crisis, other areas or regions that are not affected can come to the rescue of the affected areas. However, in a situation whereby all countries face a pandemic such as COVID-19, international cooperation often fails. Hence, the survival of each country depends on the degree of its resilience and resourcefulness. Therefore, the pandemic is a wake-up call to sub-Saharan African governments to realize the risk involved in unhealthy international agreements.

The COVID-19 pandemic has opened our eyes to three areas of self-reliance so as to mitigate the adverse effects of any global crisis. These are namely security, food and medical. In the economic development literature, the basic needs of an individual consist of food, shelter, and clothing. At the national level, the COVID-19 experience has revealed that the basic needs of a nation are security, food, and medical. While all other sectors were locked down and their activities suspended, these three sectors could not be on lockdown. This finding implies that the supply capacity in these three areas is crucial to a nation's life. Sub-Saharan African countries have a lesson to learn from this pandemic about the need to develop security infrastructure, food supply capacity, and medical facilities.

**The Way Forward**

Sub-Saharan African countries should expect a new world economic order after the COVID-19 pandemic. In other words, history shows that after a global crisis like COVID-19, there is always a change in the existing world economic order. Therefore, as the world battles with the COVID-19 pandemic, SSA countries should start bracing themselves for policy interventions that will mitigate the negative impact of the pandemic on their economies. The region should know that the new world economic order that will follow the COVID-19 pandemic may alter the existing international cooperation. This situation implies that the African sub-region should depend less on international support in her effort to alleviate the adverse effects of the COVID-19 pandemic on her economy.

From the historical review, it is observed that SSA always seeks momentary relief in the wake of a crisis. However, available evidence has shown that temporary solutions always leave their economies with prolonged economic setbacks. The focus of the region's governments should not be mainly on securing IMF and World Bank loans to cushion the effects of the pandemic on the less privileged in the society. Rather, governments in the sub-region should pay specific attention to policies that prevent long-term damage to the economy and also boost the overall economic progress of the citizens. If the temporary relief agendas like free feeding, unemployment benefits, and health benefits are the limit of the government response to the impact of the pandemic, then the region's economy will remain susceptible to future crises. Governments in the sub-region must institute policies that address the structure of the economy.

Another fact from the historical briefing is that the developed and industrialized countries' recovery process is usually faster than that of the developing countries. The developed countries have structural, institutional, and social overhead buffers to absorb shocks, unlike the developing countries that are generally weak and vulnerable. Therefore, the SSA countries should lean towards sustainability-oriented policies and programmes, such as investment in technological innovation and development of home-grown technologies, proactive support for the adaptation and diffusion of green technologies, and industrialization and manufacturing output expansion and diversification. The IMF (2020) advocates that policymakers should utilize the COVID-19 pandemic opportunity to carry out fundamental restructuring and repositioning of their economies. Such restructuring will assist in preparing for future shocks in the world economy.

The developed countries and SSA countries need different strategies for recovery from the pandemic. Put differently, SSA countries, being Third World countries, cannot imitate the First and the Second World countries in their approaches and responses to the effects of the pandemic on their economies. Each individual country has its peculiar situations and it behooves each country to identify its challenges, strengths, and prospects in designing policies to address current and future consequences of the pandemic. According to Shafaeddin (1998), since each country has specific conditions each one would require specific policies at any particular period.

Sub-Saharan African countries should not act in a hurry in their responses to the economic effects of COVID-19. Policies and decisions that are made in haste may not yield the desired results. The governments of SSA countries should patiently carry out broad consultation regarding the appropriate policy measures that will produce lasting solutions for their economies. The import substitution policy response to the post-war economic crisis in the 1960s was not appropriate. The export promotion agenda that was launched after the economic recessions of the early 1980s was also not the right policy intervention. The current leapfrogging to the service-led economic system at a time when the industrialized nations are maximizing manufacturing value-added and market share in the global market equally leaves much to be desired. Therefore, governments in the SSA sub-region will need well thought-out, home-grown policies and measures that will address both current and future crises in their economies.

**Conclusion**

This paper preempts the likely economic aftermath of the current pandemic based on historical facts. It offers guidelines on how sub-Saharan African (SSA) countries can handle their post-pandemic economic recovery plans. The COVID-19 pandemic is the world's worst global crisis since the Great Depression of the 1930s. The outbreak of the pandemic has shown the weaknesses and fragility of the SSA economies.

The scenario answers the old saying that the ultimate measure of a person, organization, or nation is not where they are in times of comfort and pleasure but their position in times of calamity. The present pandemic has revealed the true economic worth of the SSA countries. Generally, countries with deplorable medical sectors have seen the reality about their medical infrastructures. Also, those who are highly import-dependent have realized the great danger associated with such overdependence. The pandemic has revealed the weaknesses and vulnerability of SSA economies to shocks. It is imperative countries in the sub-region implement policies and programmes that will help build resilience against probable future global shocks.

This paper observes that the wrong policy interventions in addressing past crises are responsible in part for the economic backwardness of the region. The past policy responses to economic crises did not take proper cognizance of the weaknesses and strengths of the countries in SSA. In other words, the region leapfrogged prematurely in succession to import substitution, export promotion, and global value chains. This finding implies that in response to shocks from the COVID-19 pandemic, the region should adopt a gradual approach by first addressing relevant areas of sustainability in the economies of the sub-region. These will include investing in technological innovation to develop home-grown technology, upgrading essential infrastructural facilities, developing security infrastructure, and promoting economic diversification. The implementation of these policies will assist in developing a solid base for inclusive and sustainable industrialization and increased manufacturing output in the sub-region. Moreover, the high dependence of the sub-region on the developed economies will be reduced.